%% file: ann.tex
\documentclass{sig-alternate}

\usepackage[utf8]{inputenc}
\usepackage[T1]{fontenc}

\usepackage{url}
\usepackage[usenames, dvipsnames]{color}
\usepackage{listings}
\usepackage[scaled=.75]{beramono}

\usepackage{enumitem}

\definecolor{javared}{rgb}{0.6,0,0} 
\definecolor{javagreen}{rgb}{0.25,0.5,0.35} 
\definecolor{javapurple}{rgb}{0.5,0,0.35} 
\definecolor{javadocblue}{rgb}{0.25,0.35,0.75} 

\lstdefinestyle{JavaColors}
{
	captionpos=b,
	basicstyle=\ttfamily\scriptsize,
	keywordstyle=\color{javapurple}\bfseries,
       xleftmargin=15pt,
	stringstyle=\color{javared},
	commentstyle=\color{javagreen},
       numberstyle=\tiny,
	morecomment=[s][\color{javadocblue}]{/**}{*/},
      numbers=left,
	nolol=true,
	tabsize=4,
	showspaces=false,
	showstringspaces=false
      numberbychapter=false,
	numbersep=5pt,
      columns=[l]fullflexible,
}

\lstdefinelanguage{Annotation}
{
	morekeywords={public, private, protected, package, 
				int, long, short, char, byte, boolean, float, double, String, Class,
				extends, require, forbid, or, at, and, annotation,
				enum, interface, method, constructor, field, class,
				static, final, abstract,
				runtime, classfile, source},
	sensitive=true,
	morecomment=[l]{//},
	morecomment=[s]{/*}{*/},
	morestring=[b]",
}

\lstdefinelanguage{EBNF}
{
	morekeywords={Statement, Forbid, Require, TargetType, ID, ShortAtt, FloatAtt, INT, FLOAT},
	sensitive=true,
	morecomment=[l]{//},
	morecomment=[s]{/*}{*/},
	morestring=[b]",
}

\lstdefinestyle{EBNF}
{
	captionpos=b,
	basicstyle=\ttfamily\scriptsize,
	keywordstyle=\itshape,
  xleftmargin=15pt,
	stringstyle=\bfseries,
	commentstyle=\color{javagreen},
       numberstyle=\tiny,
	morecomment=[s][\color{javadocblue}]{/**}{*/},
      numbers=left,
	nolol=true,
	tabsize=4,
	showspaces=false,
	showstringspaces=false
      numberbychapter=false,
	numbersep=5pt,
      columns=[l]fullflexible,
}

\begin{document}

\conferenceinfo{SAC'15}{April 13-17, 2015, Salamanca, Spain.}

\title{A modelling language for the effective design of {\ttlit Java} annotations}

\numberofauthors{2} 
\author{
\alignauthor
Irene Córdoba\\
       \affaddr{Technical University of Madrid}\\
       \email{irene.cordoba.sanchez@alumnos.upm.es}
\alignauthor
Juan de Lara\\
      \affaddr{Autonomous University of Madrid}\\
       \email{Juan.deLara@uam.es}
}
\date{30 July 2014}

\maketitle
\begin{abstract}
This paper describes a new modelling language for the effective design of Java annotations.
Since their inclusion in the 5th edition of Java, annotations have grown from a useful tool 
for the addition of meta-data to play a central role in many popular software projects. 
Usually they are conceived as sets with dependency and integrity constraints within them; 
however, the native support provided by Java for expressing this design is very limited. 

To overcome its deficiencies and make explicit the rich conceptual model which lies behind a 
set of annotations, we propose a domain-specific modelling language.

The proposal has been implemented as an Eclipse plug-in, including an editor and an
integrated code generator that synthesises annotation processors. The language
has been tested using a real set of annotations from the Java Persistence
API (JPA). It has proven to cover a greater scope with respect to other related work in different 
shared areas of application.
\end{abstract}




\section{Introduction}\label{intro}
\input{tex/1-Introduction}

\section{Java Annotations}\label{javaann}
\input{tex/2-JavaAnnotations}

\section{Overview of the approach}\label{approach}
\input{tex/3-Overview}

\section{The Ann DSL}\label{dslann}
\input{tex/4-AnnAModellingLanguage}

\section{Tool Support}\label{tool}
\input{tex/5-ToolSupport}

\section{A Real Use Case: JPA Annotations}\label{usecase}
\input{tex/6-RealCaseExampleJPAAnnotations}

\section{Related Research}\label{related}
\input{tex/7-RelatedResearch}

\section{Conclusions and Future Work}\label{conclusions}
\input{tex/8-ConclusionsAndFutureWork}

\noindent{\bf Acknowledgements.} This work was supported by the Spanish Ministry of Economy and Competitivity 
with project Go-Lite (TIN2011-24139) and the Community of Madrid with project SICOMORO (S2013/ICE-3006).

\bibliographystyle{acm}
\bibliography{ann} 

\end{document}

%% file: tex/1-Introduction.tex
In 2004 annotations where added to the Java language as an answer to the huge 
amount of boilerplate code that many APIs, such as JavaBeans or JAX Web Services (JAX-WS), required \cite{javaurl}. 
Other authors explain the appearance of annotations in Java as a result of
the increasingly growing tendency of including the meta-data associated with a program
within the program itself instead of keeping it in separate files; as well as the pressure
from other programming languages which already included similar features, like C\# \cite{Schildt:javaann}.

Since their introduction in the language, annotations have become a success and
are widely used in many important projects within the software development scene. We find them
in frameworks like Seam \cite{seam} and Spring \cite{spring}, in the Object Relation Mapping of Hibernate 
\cite{hibernate}, and also in proper Sun Java standards such as the aforementioned JAX-WS, JavaBeans,
Enterprise JavaBeans and the Java Persistence API (JPA) \cite{jpa}.

However, despite this success, the native support that Java provides for their development is very poor. 
On the one hand, the syntax for defining annotations is rather unusual for an accustomed Java programmer:
some Java constructions are reused for other purposes that absolutely differ from their usual semantics. 
On the other hand, annotations are rarely conceived in an isolated way; instead they are usually part of
a set with dependencies and integrity constraints. Moreover, each annotation separately usually carries integrity
constraints with respect to the elements it can be attached to. Currently there is no effective way in Java
for making explicit the constraints underlying a set of annotations at design time. Instead, the usual
path taken to overcome this deficiencies is to develop an extension to the Java compiler to ensure that such constraints are complied with.

As a first step towards the alleviation of this situation, we propose {\em Ann}, a Domain-Specific Language 
(DSL) \cite{Brambilla:mdse} aiming to provide a
more expressive and suitable syntactic support for the design of sets of annotations and their associated integrity constraints. 
We have developed an integrated development environment as an Eclipse plug-in.
The environment includes a code generator to translate the design and constraints expressed using {\em Ann} into Java code, which can be fully integrated in projects in such language. \emph{Ann} has been tested using a real set of annotations from JPA,
demonstrating that it can capture a wide set of the constraints in its specification. More information and source code
of the project is available at \url{https://github.com/irenecordoba/Ann}.

The rest of the paper is organised as follows: section \ref{javaann} gives a more detailed overview on the current limitations of
Java annotations; section \ref{approach} overviews our approach;
section \ref{dslann} describes the proposed DSL, {\em Ann}; section \ref{usecase} details a real case study for {\em Ann}; section
\ref{related} compares our proposal with related work; and section \ref{conclusions} summarises the conclusions and 
future development.

%% file: tex/2-JavaAnnotations.tex
To help understanding the current limitations of Java annotations, in this section we describe how 
they are defined in Java (subsection~\ref{sec:jann_def}), and how is their correct use checked (subsection~\ref{sec:ann_proc}).

\subsection{Defining Java annotations}\label{sec:jann_def}

Java annotations do not constitute a type of their own. Instead, they are defined as \emph{special}
interfaces. Listing \ref{java_def} shows an example of the definition of a simple annotation (called
an \emph{annnotation type}).

\begin{lstlisting}[
	style=JavaColors, 
	language=Java, 
	caption=Annotation \texttt{Person} defined in Java.,
	label=java_def
]
package examples;

import java.lang.annotation.Target;
import java.lang.annotation.ElementType;

@Target(ElementType.TYPE)
public @interface Person {
    String name() default "Mary";
    int age() default 21;
    float weight() default 52.3f;
}
	\end{lstlisting}

As it can be noticed, the special nature of annotations is pointed out by the \texttt{@} character 
before the \texttt{interface} keyword (line 7). The zero-argument methods inside the container (lines 8-10)
are the \emph{fields} (the parameters) of the annotation. The notation is cumbersome because Java is 
providing a syntax characteristic of one construction (a method) to specify another completely different
(an annotation parameter). Moreover, to assign a default value to those \emph{fields},
the keyword \texttt{default} must be used, instead of the equality symbol, more natural and common in this context.

Finally, line 6 shows an example of an annotation being used: {\tt Target}. This annotation is used to specify
what kind of elements the declared annotation can annotate (\emph{targets}). Although it is a way to introduce constraints in the 
definition of an annotation, it is very limited. For example, in this case by using the value {\tt TYPE} of the
enumeration {\tt ElementType}, {Person} can only be applied to classes, interfaces (including annotation types) and 
enumerations. However, there is no way to e.g., to restrict its applicability to classes only.

This was a simple example, but if we take a look at the JPA documentation, we find that the annotation \texttt{Entity} can 
only be applied to classes meeting the following more elaborated requirements  \cite{jpa}:
\begin{itemize}[noitemsep]
	\item They must have a public or protected constructor.
	\item They must not be final.
	\item They must not have any final method.
	\item Their persistent fields must be declared private, protected or package-private.
\end{itemize}

None of these statements can be expressed nowadays with the syntax available for the definition of annotations.

What is more, when designing annotation sets, it is common to have constraints involving several annotations. For
example, the JPA annotation {\tt Id} is only allowed in attributes within classes annotated with {\tt Entity}.
We call such constraints the static semantics or integrity constraints of an annotation (set). 

Therefore, what can be done to ensure the compliance of such outlined constraints? The only remaining choices are to write a 
guiding comment for its use and signal an error at runtime. In addition, it is possible to develop extensions to the 
Java compiler, known as annotation processors, an option we will detail more in the following subsection.

\subsection{Annotation processors}\label{sec:ann_proc}

The Java package \texttt{javax.annotation.processing} provides a set of elements for processing annotations at compile
time. An annotation processor is invoked by the compiler, and it can check the annotations attached to any
program element, performing an arbitrary task. Typically, the processor will check the correctness of the annotation
placement (i.e., its static semantics), and may perform further actions (e.g., generating code).
Annotation processing works in rounds. In each round a processor may be required to process a subset of the annotations
found in the source code and the binary files produced in the prior round. If a processor was executed in a given round,
it will be called again in the next rounds.

Listing \ref{ann_proc} shows the structure of a typical annotation processor. Line 1 specifies the annotation to be checked, 
{\tt Person} in this case. 
The key method of the processor is {\tt process} (lines 5-23), where the elements annotated with the particular annotation
are looked up and checked. If any of them does not satisfy the checks, then an error is raised using the functionality provided by the 
processing package (lines 15-20).

\begin{lstlisting}[
	style=JavaColors, 
	language=Java, 
	caption=Structure of an annotation processor.,
	label=ann_proc
]
@SupportedAnnotationTypes("Person") // annotation to be checked
@SupportedSourceVersion(SourceVersion.RELEASE_6)
public class PersonProcessor extends AbstractProcessor 
{ 
	@Override
	public boolean process(Set<? extends TypeElement> annotations, 
                                   RoundEnvironment objects) 
    {	
        // iterate on all objects to check
        for (Element elt: objects.getElementsAnnotatedWith(Person.class)) 
        {
            // evaluate correct placement of Person annotation for elt
            ...
            // if error
            this.processingEnv.getMessager().printMessage
            (
                Kind.ERROR, 
				"The annotation @Person is disallowed for this location.", 
                elt
            );
        }	
		return true;
	}
}
	\end{lstlisting}

It is important not to confuse annotation processing with reflection. While the former takes place at compile time, the latter
is at execution time. The values of an annotation at a given program element can be checked at execution time via the Java 
Reflection API, but it has several disadvantages, like an overhead in performance,
the requirement of runtime permission (which may not be granted), and the possibility of breaking object-oriented
abstractions. 

In the context of checking the correctness of annotations, it is more appropriate to do it via annotation processors, because
they can find and signal the errors without the need to execute the program.
However, coding such processors is tedious and error prone. Moreover, we believe it would be advantageous to make
explicit the underlying annotation constraints at a higher level, together with the annotation structure. For this purpose, we have
created {\em Ann}, a DSL to define the structure and integrity constraints of Java annotations.

%% file: tex/3-Overview.tex
\begin{figure}[h!]
\centering
	\includegraphics[width = 7.5cm]{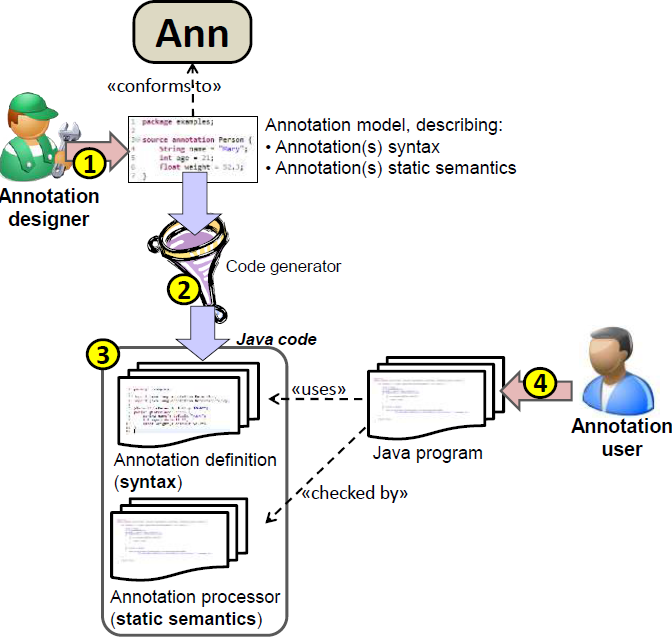}
  \caption{Overview of our approach.}
	\label{fig:overview}
\end{figure}

Figure~\ref{fig:overview} shows the working scheme of our approach to solve the
problems outlined in section \ref{javaann}. The main idea is to use
a DSL~\cite{Voelter} called Ann, to describe the syntax and static semantics
of the family of annotations to be built in a declarative way (label 1 in the scheme). 
This DSL provides appropriate
primitives for this task, beyond those natively offered by Java. 

Our solution includes a code generator (label 2) that produces plain Java files with 
the syntax definition and the annotation processors for the defined annotation (label 3). Then, the
annotations can be safely used, because their correct use in Java programs is checked by the
generated annotation processors. 

Altogether, using Ann has several advantages, including: (i) it allows to make explicit the structure and integrity constraints
of a set of annotations in a high-level, declarative way; and (ii) it automatically produces the annotation
processors to check the correct use of annotations. 

The next section details the elements of the Ann language.

%% file: tex/4-AnnAModellingLanguage.tex
As we have previously stated, Ann is a domain-specific modelling language aimed at the description of 
the syntax and static semantics of Java annotations. Modelling languages are conceptual tools for 
describing reality explicitly, from a certain level of abstraction and
under a certain point of view \cite{Brambilla:mdse}. They are defined by three key elements: abstract syntax, concrete syntax and 
semantics.

The next three subsections describe the abstract, concrete syntax and semantics of Ann.

\subsection{Abstract syntax}

\begin{figure}[h!]
\centering
	\includegraphics[width = 6.4cm]{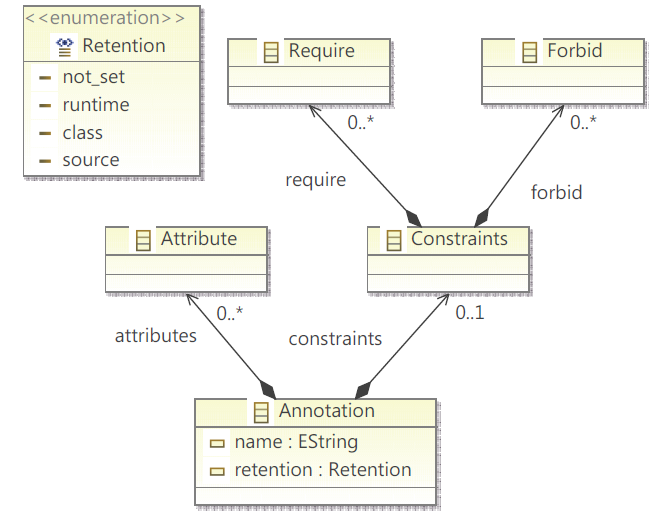}
  \caption{Simplified meta-model excerpt representing the abstract syntax of Ann.}
	\label{metam}
\end{figure}

The abstract syntax describes the structure of the language and the way its different elements
	can be combined. It has been specified in Ann by using a meta-model\footnote{A meta-model is a model which describes
	the properties of a set of models, and so it is in a higher level of abstraction}, which can be found simplified in 
	Figure \ref{metam}. 

\begin{figure}[h!]
\centering
	\includegraphics[width = 6.2cm]{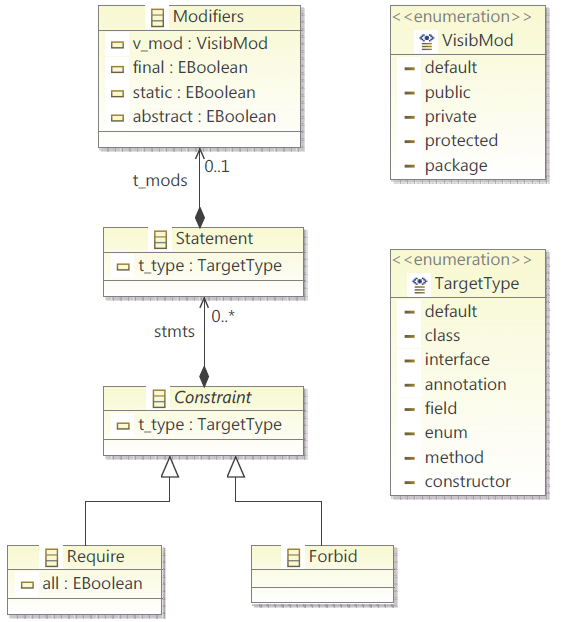}
  \caption{Meta-model excerpt for annotation constraints.}
	\label{metam-const}
\end{figure}

The {\tt Annotation} meta-class contains both the attributes of an annotation and its associated constraints.
Details concerning attributes have been omitted; and constraints are split into two types: requirements 
(class {\tt Require}) and prohibitions (class {\tt Forbid}). 
In Figure \ref{metam-const} we can see an expanded
section of this meta-model, in particular the one concerning the constraints.
		
Each \emph{statement} represents a description of a Java element (like {\tt class}, {\tt interface} or {\tt field}) 
over which the annotation is (dis-)allowed. 
Several statements are possible within the same constraint (e.g., if the same annotation can be applied to several targets), 
enhancing the expressive power of Ann. 
There is also the possibility of expressing constraints for specific target types (e.g., a {\tt field}), 
which indicates that the given constraint only applies when the annotation is attached to that target type (e.g., a {\tt field}). 
An annotation is correctly placed at a target type if it satisfies some of the statements of the positive 
requirements for the given target, and none of the prohibitions.

It will be shown in Section \ref{usecase} that these
	two types of constraints, and their combinations have enough expressive power to cover a huge scope of 
	the full conceptual model of an annotation group design in a real use case.

\subsection{Concrete syntax}
The concrete syntax of a DSL describes the specific representation of the language, and hence how users
visualize or create models. Concrete syntaxes can be textual or graphical. 
Given that one of the goals of Ann
is to give a friendlier syntax for Java developers defining annotations, mitigating the incoherences that can be found
nowadays in Java language, a textual concrete syntax has been chosen for it.

An excerpt of the concrete syntax definition for the constraints within an annotation can be found in Listing \ref{sint_c},
represented in Extended Backus-Naur Form.

\begin{lstlisting}[
	style=EBNF, 
	language=EBNF, 
	caption=Concrete syntax excerpt for constraints in Ann.,
	label=sint_c
]
<Forbid> ::= 	
	"forbid" <Statement> ("and" <Statement>)* ";" |
  "at" <TargetType> ":" 
		"forbid" <Statement> ("and" <Statement>)* ";";

<Require> ::= 
	"require" <Statement> ("or" <Statement>)* ";" |
	"at" <TargetType> ":" 
		"require" "all"? <Statement> ("or" <Statement>)* ";";

	\end{lstlisting}

Listing \ref{ann_def} shows how the Java annotation type \texttt{Person} previously shown in Listing \ref{java_def}
would be described using Ann. 
A new keyword (\texttt{annotation}) is used on its declaration (line 3), and the overall result 
is a clearer, more readable code. The 
definition and initialisation of the attributes is now in harmony with the usual Java syntax 
for the same purpose. 

\begin{lstlisting}[
	style=JavaColors, 
	language=Annotation, 
	caption=Annotation \texttt{Person} defined in Ann.,
	label=ann_def
]
package examples;

annotation Person {
    String name = "Mary";	
    int age = 21;
    float weight = 52.3;
		
    require public class;          // annotation allowed for classes...
		
    at class: forbid final field;  // ... with no final fields
}
\end{lstlisting}

Regarding the restriction of the allowed targets, we can now express some
more elaborated descriptions, in this case that Person can only annotate public classes (line 8)
with no final fields (line 10).
We recall that with Java the closer we got to this statement was that the annotation could have as targets
classes, interfaces and enumerations, which is much more general than we intend.

In the concrete syntax for requirements, we note also the special keyword {\tt all}. This would apply if, for instance, we would want
that all the methods of the classes annotated with {\tt Person} were also public. Then we would add the clause {\tt at class: require
all public method}.

\subsection{Semantics: code generation}
The semantics of a modelling language can be specified by several means, like e.g., providing an interpreter or a code generator. In the 
present case, code generation has been the adopted solution. 

In order to fully specify the semantics of Ann, it is necessary to generate on one hand the Java code associated with the definition of
the annotations; and on the other the code of the processors. The latter will ensure that the constraints specified for 
each of the defined annotations are being met.

For each of the annotations defined at most two processors will be generated, one for checking the requirements and the other for 
checking the prohibitions. The structure of the annotation processors generated complies with the one presented in Section \ref{javaann}:
each of the relevant elements of the Java program is looked up to check whether its properties satisfy the specified 
requirements or prohibitions.

%% file: tex/5-ToolSupport.tex
Ann has been developed using the Eclipse Modelling Framework (EMF)~\cite{EMF}.
Integrated in this environment, different tools have been used for the different elements of the DSL.
The meta-model has been described using the meta-modelling language Ecore, which is based in a subset
of UML class diagrams for the description of structural aspects.

\begin{figure}
\centering
	\includegraphics[width = 7cm]{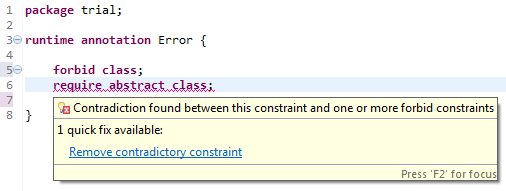}
  \caption{Validation of contradictory constraints.}
	\label{qf}
\end{figure}

Xtext \cite{xtext} has been used to define the textual concrete syntax. Xtext is integrated with EMF 
and able to generate a fully customisable and complete editor for the defined language. In our present case we have added the 
validation of, among other issues, contradictory constraints specified within an annotation, providing
relevant quick fixes, as can be seen in Figure \ref{qf}.

Finally, the code generator has been developed using the language Xtend, included in the framework Xtext. 
Xtend is a Java dialect more expressive and flexible, with facilities for model navigation. It also
allows creating generation templates, what makes it specially useful for code generation.

The result is an Eclipse plug-in, which is seamlessly integrated within the Eclipse Java Development
Tools (JDT).

%% file: tex/6-RealCaseExampleJPAAnnotations.tex
A subset of JPA annotations has been chosen in order to test the Ann DSL:
{\tt Entity}, {\tt Id}, {\tt IdClass}, {\tt Embeddable} and {\tt EmbeddedId}. 

This selection has been made according to their extensive use in the JPA context, given that all of them
are used to describe entities and their primary keys, central concepts in database design.

\subsection{Defining the annotations with Ann}
The constraints associated with the {\tt Entity} annotation were outlined in Section \ref{javaann}. Moreover, 
given that it defines an entity within a database, a corresponding primary key must also be specified. The other
selected annotations are used precisely for this purpose.

An entity may have a simple or compound primary key. For the former case, the annotation {\tt Id} is used; for the latter,
the annotations {\tt IdClass} or {\tt EmbeddedId} and {\tt Embeddable}\footnote{Depending on whether using fields or an embeddable class to represent the compound key, respectively.}. 

	\begin{lstlisting}[style=JavaColors, language=Annotation, caption=Selected JPA annotations defined in Ann., label=jpa_cod]
runtime annotation Entity {
	String name = "";

	require class;
	forbid final class;
	
	at class: require public constructor or protected constructor;
	at class: forbid final method;
	
	at class: require @Id method or @Id field or 
                      @EmbeddedId method or @EmbeddedId field;
	at class: forbid @Id method and @EmbeddedId method;
	at class: forbid @Id field and @EmbeddedId field;
}

runtime annotation Embeddable {
	require class;
	
	at class: forbid @Id method;
	at class: forbid @EmbeddedId method;
	
	at class: forbid @Id field;
	at class: forbid @EmbeddedId field;
}

runtime annotation EmbeddedId {
	require method or field;
	
	at field: require @Entity class;
	at method: require @Entity class;
}

runtime annotation Id {
	require method or field;
	
	at field: require @Entity class;
	at method: require @Entity class;
}

runtime annotation IdClass {
	Class value;
	
	require @Entity class;
}
	\end{lstlisting}

Listing \ref{jpa_cod} shows the description of the explained annotations using Ann. Clearly 
the chosen subset of annotations is very interrelated given all the respective constraints that can be noticed. 
For example, a class annotated with {\tt Embeddable} (lines 16-24) acts as a primary key for another class, in which it is embedded, 
and thus it must not have a primary key itself, prohibition which is expressed through lines 19-23. 

Alternatively, the annotation {\tt IdClass} (lines 40-44) can be used to specify the class that contains 
the fields which form the compound primary key. Therefore it can only be attached to classes annotated with {\tt Entity},
requirement described in line 43. 

Annotations {\tt Id} (lines 33-38) and {\tt EmbeddedId} (lines 25-30) mark the primary key of an entity, and thus can only
annotate methods or fields (lines 34 and 27 resp.) which form part of a class annotated with {\tt Entity} (lines 36-37 and
29-30 resp.).

Finally, regarding the {\tt Entity} annotation (lines 1-14), structural properties of the annotated classes are expressed throughout
lines 4-8; and lines 10-11 establish the need
of a primary key through a requirement, among other constraints.

After the definition of all the annotations and their constraints, the corresponding code is generated and ready to
use in both new or existing Java projects.

\subsection{Using the generated code}
The generated processors are capable of detecting where a constraint is being violated and also notify the developer by means
of an explanatory message. 

\begin{figure}
\centering
	\includegraphics[width = 8cm]{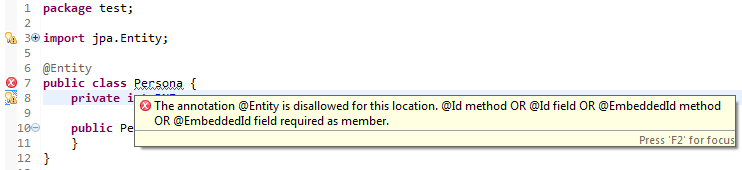}
  \caption{Entity without primary key.}
	\label{err1}
\end{figure}

In Figure \ref{err1} the annotation {\tt Entity} is being used on a class and no primary key is being specified, situation not allowed
in the JPA context.

\begin{figure}
\centering
	\includegraphics[width = 8cm]{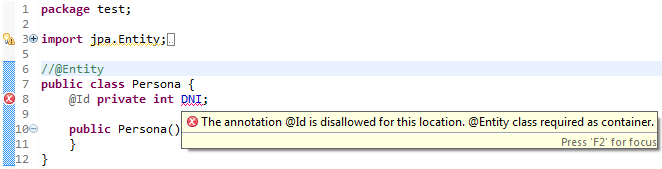}
  \caption{Primary key in a field not belonging to an entity.}
	\label{err2}
\end{figure}

Another example of misuse is the one shown in Figure \ref{err2}. In this case, the annotation {\tt Id} is used in a field inside 
a class that is not annotated as {\tt Entity}, situation that leads to another error.

%% file: tex/7-RelatedResearch.tex
Some research has been made in order to improve and expand the functionality of Java annotations.
For example,  Phillips in~\cite{Phillips:2009:CMA:1639950.1640005} aims at conciliating object oriented principles
with the design of annotations by the introduction of a new one: {\tt composite}. With it, he manages to
support composition, allowing encapsulation and polymorphism of annotations.

A Java extension, {\tt @Java}, is proposed by Cazzola and Vacchi~\cite{Cazzola20142}
in order to expand the range of application of an annotation to code blocks and expressions, although some
improvement in this respect has also been made natively in the latest version of Java \cite{java8}.

The expressiveness limitations of Java annotations are recognised in~\cite{ClarkSW08}, where a proposal
is made to embed DSLs into Java, with a more natural and flexible syntax. JUMP~\cite{JUMP} is a tool
to reverse engineer Java programs (with annotations) into profiled UML class diagrams.

Although the aforementioned approaches expand the features of Java annotations, they do not address the 
integrity and design issues explained throughout this paper, which is the main goal of our work. 

Just a few works are aimed at improving the design of annotations.
Darwin~\cite{Darwin:annabot} suggests a DSL, called AnnaBot, based on \emph{claims} about 
a set of existing annotations, with a concrete syntax very similar to Java. With this \emph{claims} interdependency
constraints can be expressed within a set of annotations. However, there is no possibility of characterising the
targets of an annotation type. Moreover, no improvement is made with respect to the syntax for defining 
annotations in Java, given 
its heavy focus on existing sets of annotations and constraints between them, and not on isolated ones. Finally,
the approach uses reflection to check the statements of its \emph{claims}, which could and should be avoided.

Another approach is AVal~\cite{noguera:aval}, a set of meta-annotations\footnote{Annotations whose 
targets are other annotations.} to add integrity constraints at the definition of the annotation type. 
This approach has as a drawback that its expressive possibilities are rather restricted, given the limited 
flexibility which meta-annotations provide. For example, a simple constraint we saw for entities in Section \ref{javaann}
was that no class methods should be final, and this cannot be expressed by the meta-annotations provided
in AVal.

Hence, to the best of our knowledge, Ann is the first proposal for a high-level means to describe both the
syntax and well-formedness constraints of annotation families, making explicit the design of such annotation set
and allowing their immediate use on Java projects thanks to the code generation facility.

%% file: tex/8-ConclusionsAndFutureWork.tex
Ann makes possible the effective design of Java annotations by improving their native syntactical support and
allowing the expression of integrity constraints both related to an annotation type and within a set of annotations.
Thanks to the code generator, the approach can be perfectly integrated with existing Java projects. 
Moreover, with the use of annotation processors all the integrity constraints described with the DSL are 
checked at compile time, which improves both usability and efficiency. This is because it is not necessary to
execute the application in order to know whether the annotations are being correctly used, 
hence saving much time and effort for developers.
 
Concerning future work, a huge range of possibilities is available given the flexibility
that a DSL provides. As seen in Section \ref{related}, the meta-model of Java annotations can be still
improved and expanded to improve its harmony with the rest of Java elements, like, for example, 
its conciliation with object-oriented principles such as composition, inheritance and polymorphism,
which might help to make cleaner the design of a set of annotations.

At present two basic types of constraints are considered in Ann (requirements and prohibitions), which
are enough to express common integrity constraints as it has been seen in Section \ref{usecase}. However, 
further experimentation could reveal new constraint types or combinations, which could be added 
to the DSL in the future, given the flexibility that a meta-model provides. Another line of work is the reverse engineering of annotation constraints from the analysis of annotated Java programs.